\documentclass{article}
\usepackage{spconf,amsmath,graphicx}

\usepackage{mathtools}
\usepackage{amssymb}
\usepackage{booktabs}
 
\usepackage{multirow}
\usepackage{enumitem}
\usepackage{multicol}
\usepackage{color, colortbl}
\usepackage{xcolor}
\usepackage{comment}
\usepackage{cite}
\usepackage{subcaption}
\usepackage{threeparttable}
\usepackage{hyperref}
\usepackage{tikz}
\usepackage{tabularx}

\colorlet{full}{green!10}
\colorlet{cmn}{teal!10}
\colorlet{FOMAML}{purple!10}
\colorlet{MTL}{purple!10}
\colorlet{MTL*}{purple!10}
\colorlet{multi}{purple!10}
\colorlet{euro}{orange!10}

\name{Shih-Heng Wang$^{1}$, Jiatong Shi$^{2}$, Chien-yu Huang$^{1}$, Shinji Watanabe$^{2}$, Hung-yi Lee$^{1}$ }
\address{National Taiwan University$^{1}$, Carnegie Mellon University$^{2}$}

\ninept

\begin{document}
\title{Fusion of Discrete Representations and Self-Augmented Representations for Multilingual Automatic Speech Recognition}
\maketitle

\begin{abstract}
Self-supervised learning (SSL) models have shown exceptional capabilities across various speech-processing tasks. Continuous SSL representations are effective but suffer from high computational and storage demands. On the other hand, discrete SSL representations, although with degraded performance, reduce transmission and storage costs, and improve input sequence efficiency through de-duplication and subword-modeling. To boost the performance of discrete representations for ASR, we introduce a novel fusion mechanism that integrates two discrete representations. The fusion mechanism preserves all the benefits of discrete representation while enhancing the model's performance by integrating complementary information. Additionally, we explore ``self-augmented'' discrete representations, which apply transformations to a single continuous SSL representation, eliminating the fusion mechanism's dependency on multiple SSL models and further decreasing its inference costs. Experimental results on benchmarks, including LibriSpeech and ML-SUPERB, indicate up to 19\% and 24\% relative character error rate improvement compared with the non-fusion baseline, validating the effectiveness of our proposed methods.

\end{abstract}


\section{Introduction}
\label{sec: intro}

Self-supervised learning (SSL) models have demonstrated exceptional success across a variety of speech-processing tasks~\cite{hsu2021hubert, yadav-sitaram-2022-survey, 9023195,baevski2020wav2vec,  babu22_interspeech, chung2021w2v, Chen2021WavLMLS, Mohamed_2022, shi2024multiresolution,pratap2023mms, yang21c_interspeech}.
Prior works mostly focused on leveraging continuous SSL representations~\cite{yang21c_interspeech, shi2023mlsuperb}, which, despite their effectiveness, are notorious for their high storage and computational costs. To address these issues, recent researches ~\cite{chang2023exploring, chang22e_interspeech, UniversalSpeechDiscreteTokens, AcousticBPE, Voxtlm, erdogan23_interspeech, 10387745, kim2023lip} have shifted towards discrete SSL representations. These representations, obtained by applying discretization to continuous SSL features, offer significant benefits such as lower transmission, storage costs, and faster I/O during training (see Section~\ref{subsec:statistics}). Specifically, in the context of automatic speech recognition (ASR), using discrete representations brings an additional advantage: it significantly reduces the input sequence length without substantially affecting performance. For instance, ~\cite{chang23b_interspeech} have shown de-duplication and Byte Pair Encoding (BPE) subword-modeling reduce the sequence length to nearly one-third of its original size without severe performance degradation.

While using discrete representations for ASR offers the above-mentioned advantages in computation and storage, when compared to continuous representations, their performance often falls short. Hence, enhancing the performance of discrete representations in ASR while keeping the computational costs manageable presents a non-trivial challenge. Motivated by the progress in continuous SSL representations~\cite{srivastava2023effuse, FeaRLESS, Berrebbi2022CombiningSA}, we find offering the model with complementary information would be a straightforward way to boost performance. For instance, ~\cite{Berrebbi2022CombiningSA} proposed to fuse SSL features with spectral features to provide the model with domain-robust information. 
~\cite{FeaRLESS} explored ways to effectively fuse multiple continuous SSL features to enhance performance. Considering the enormous improvement brought by leveraging multiple features, we aim to investigate ways to leverage multiple discrete representations to improve ASR performance.



\begin{figure}[t]
    \centering
    \includegraphics[width=0.93\linewidth]{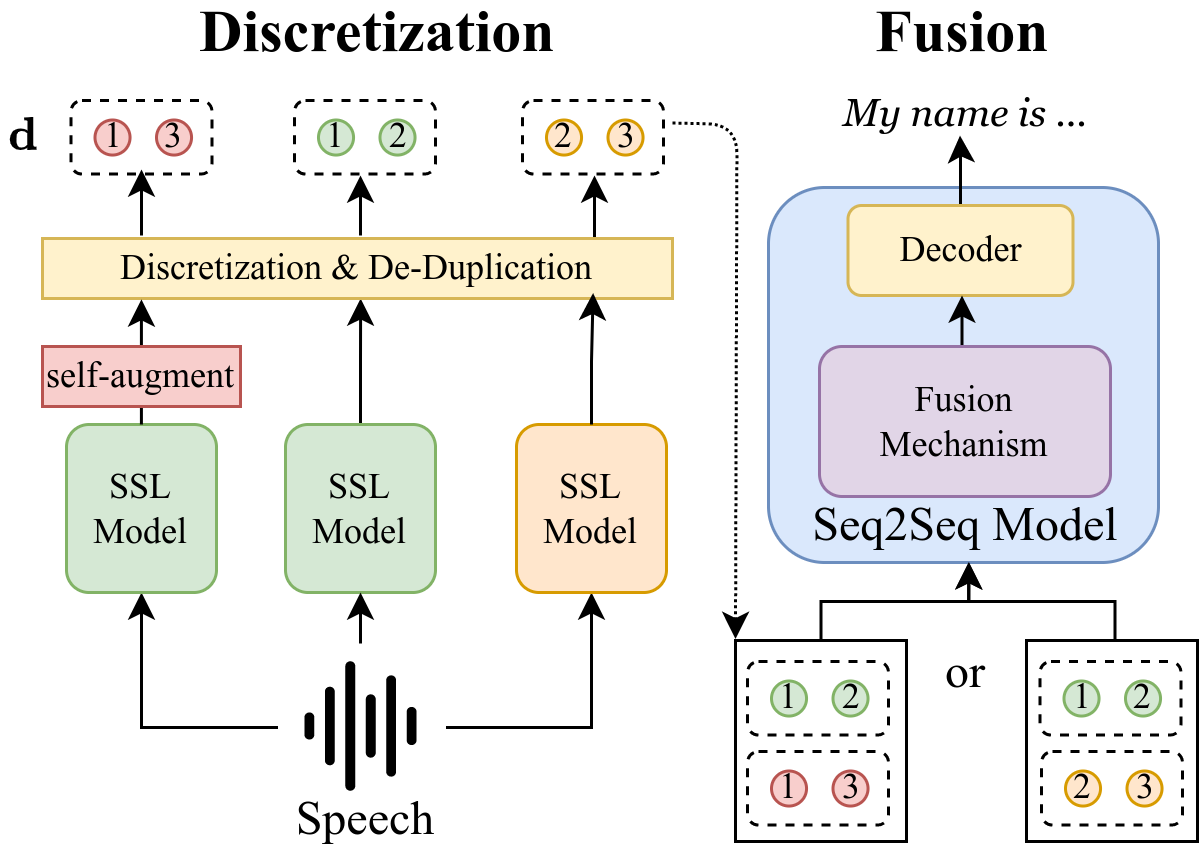}
    \caption{General pipeline of our proposed fusion mechanism.}
    \label{fig:general}
\end{figure}
In this paper, we explore the fusion of discrete representations for ASR to provide the model with complementary information. Figure~\ref{fig:general} offers a high-level overview of our fusion pipeline, including the discretization process and subsequent fusion. Through the discretization process (see Section~\ref{subsec:discretization}), we derive the final discrete representations $\mathbf{d}$. Then, we apply our proposed fusion mechanism to utilize two such discrete representations simultaneously, enhancing the model’s performance by integrating complementary information. Existing studies~\cite{kim2023manytomany, choi23_interspeech, kim2024tmt, MultiModalTokens}, which share a similar concept with discrete representation fusion, are primarily within the multi-modal domain. To leverage discrete representations from multiple modalities, they typically standardize each of their frame rates to solve the linear misalignment between representations. However, this method conflicts with sequence length reduction techniques like de-duplication and BPE subword modeling. De-duplication condenses consecutive identical tokens into a single token, and BPE subword modeling combines frequent patterns of tokens into a new token. Both of them will lead to non-linear misalignment between representations and complicate straightforward fusion. On the other hand, concatenating representations along the time dimension will lead to lengthy input sequences, diminishing the benefits of using discrete representations for ASR.

To address these challenges, we propose a novel fusion mechanism that integrates two non-linearly misaligned discrete representations. Our approach leverages an attention-based mechanism to intelligently learn the alignment between representations. To the best of our knowledge, this is the first investigation into such a fusion approach specifically for discrete representation-based ASR, aiming to preserve the benefits of discrete representations while enhancing ASR performance.
Additionally, we investigate ``self-augmented'' discrete representations (see Figure~\ref{fig:general}), which are created by applying simple and efficient transformations to a single continuous SSL representation. Using these ``self-augmented'' representations in our fusion mechanism helps eliminate the dependency on the second SSL model, thus broadening the applicability of our approach and reducing inference costs (see Section~\ref{subsec:analysis}). Moreover, our analysis indicates that employing fusion with self-augmented discrete representations yields more robust results than using fusion with another SSL discrete representation (see Section~\ref{subsec:analysis}).

To validate our approach, we conducted evaluations on two well-known benchmarks: LibriSpeech~\cite{libirspeech} and ML-SUPERB~\cite{shi2023mlsuperb}, following the protocols of the \textit{Interspeech2024-Discrete-Speech-Unit-Challenge}~\cite{chang2024interspeech}. The experimental results demonstrate that our fusion mechanism consistently improves the character error rate (CER) across various test sets. Notably, the fusion of two different SSL features resulted in the most significant improvements, achieving a 19\% and 24\% relative reduction in CER for the benchmarks, respectively. Fusion with self-augmented representations also yielded at most 6\% and 19\% relative reductions in CER for each benchmark dataset. Combining the above findings, our submission to the \textit{Interspeech2024-Discrete-Speech-Unit-Challenge}~\cite{chang2024interspeech} achieved the lowest CER among all submissions, and an overall second place considering the bitrate ranking. 
\section{Discrete Representation for ASR}
\subsection{Discretization process}
\label{subsec:discretization}
\begin{figure}[t]
    \centering

    \begin{subfigure}[h]{0.5\linewidth}
    
    \includegraphics[width=\linewidth]{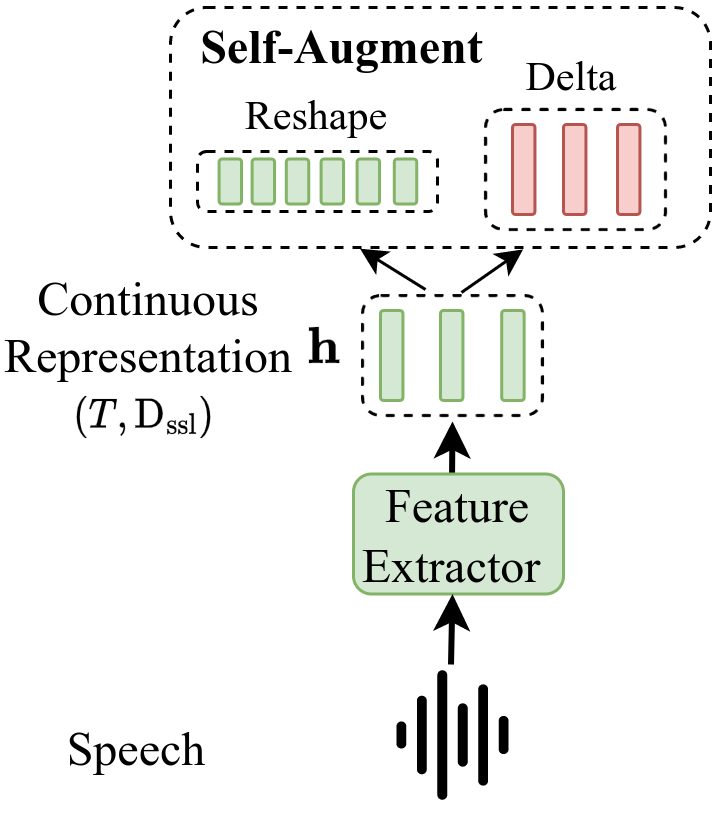}
    \caption{Derive continuous representation.}
    \label{fig:sub1}
  \end{subfigure}
  \begin{subfigure}[h]{0.45\linewidth}
    
    \includegraphics[width=\linewidth]{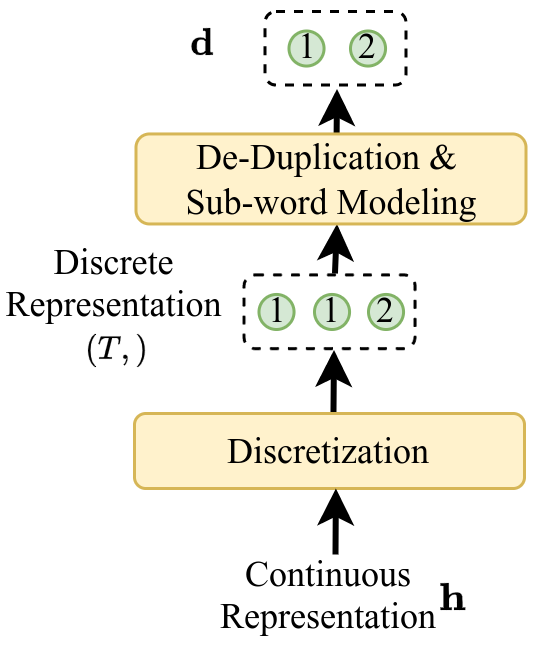}

    \caption{Derive discrete representation}
        \label{fig:sub2}
  \end{subfigure}
  
      \caption{The pipeline for extracting discrete representations.}
  
    \label{fig:discretization}
\end{figure}

Figure~\ref{fig:general} provides a high-level overview of our fusion pipeline. In this section, we would like to introduce how to derive the final discrete representations $\mathbf{d}$ for fusion. Figure~\ref{fig:discretization} shows the process of discretization of speech inputs (the Self Augment part will be introduced in Section~\ref{sec:repr_tfm}).
First, in Figure~\ref{fig:sub1}, a feature extractor processes the raw waveform input to obtain $T$ continuous representations $\mathbf{h} = [h_1, h_2, ..., h_T]$, $h_i \in \mathbb{R}^{\text{D}_\text{ssl}}$, where $\text{D}_\text{ssl}$ indicates the output dimension of the feature extractor.
Then, in Figure~\ref{fig:sub2}, to convert continuous representations into discrete ones, we could apply quantization methods like vector quantization or K-means clustering over all frames.
Each frame of continuous representation is converted to a discrete unit (represented as integers of circles in Figure~\ref{fig:general} \& \ref{fig:sub2}), resulting in a sequence of discrete units, referred to as discrete representation.
Considering the speech characteristics, the obtained discrete representation is generally quite long. For example, employing HuBERT~\cite{hsu2021hubert}, a famous SSL model, as the feature extractor will introduce 50 discrete units per second. Hence, we follow ~\cite{chang2023exploring} and apply de-duplication to remove consecutive repeating tokens. Then, subword modeling is used to further reduce the sequence length, yielding the final discrete representation $\mathbf{d}$.

\subsection{Storage, I/O and length reduction statistics}
\label{subsec:statistics}
In this section, we provide detailed statistics about the storage requirements and input sequence length for discrete representations in ASR. We have gathered these statistics from our Train \& Dev set, which utilizes MMS-1B~\cite{pratap2023mms} as the feature extractor. Further details about our Train \& Dev set are available in Section~\ref{subsec:dataset}.

\noindent \textbf{Storage \& I/O.}
\begin{table}[t]
    \centering
     \caption{Required storage space comparison between continuous and discrete representations. The reduced ratio indicates that the discrete representations take only 0.04\% of the space required by their continuous counterparts.}
\begin{tabular}{l|c|c|c}
\toprule
\textbf{Splits} & \textbf{Continuous} & \textbf{Discrete}  & \textbf{Space Ratio} \\ \midrule
Train            & 290.85 GB            & 121.43 MB  &  0.04\%\\ \midrule
Dev              & 47.43  GB          & 19.44 MB  &  0.04\% \\ 
\bottomrule
\end{tabular}

    \label{tab:storage}
\end{table}
As discussed in Section~\ref{sec: intro}, using discrete representations for ASR offers advantages such as low storage costs and faster I/O during training. To provide a clearer picture, we present statistics on the storage space required for discrete representations, enabling a quantitative comparison with continuous representations. According to Table~\ref{tab:storage}, discrete representations occupy only 0.04\% of the space of their continuous counterparts. Specifically, our training set requires only 121 MB of storage, which can easily be loaded into RAM for faster I/O during training. In contrast, loading the entire training set with continuous representations is challenging and costly. Consequently, training the ASR model with discrete representations is faster than with their continuous counterparts. 

Moreover, the enhanced I/O speed also benefits the transmission bitrate. Transmitting continuous representations requires approximately 2M bits per second\footnote{Calculated as 50 (unit/sec) * 1,280 (dim) * 32 (fp32)\label{fn:2M}}, whereas discrete representations significantly lower the transmission bitrate to less than 300 bits per second. We will further detail the bitrate of our proposed methods in Section~\ref{sec:quantitative_result}. This comparison underscores the efficiency of using discrete representations, not only in storage and ASR training but also in data transmission.

\noindent \textbf{Length reduction.}
As demonstrated in \cite{chang23b_interspeech}, the application of de-duplication and BPE subword-modeling can significantly reduce the input sequence length for discrete representation ASR. To illustrate this effect, we present statistics on the input sequence length of our discrete representations in Table~\ref{tab:sequence}. 
This table shows how these length-reduction techniques impact the sequence length, both with and without their application. Our results indicate that sequence length can be halved\footnote{We set the BPE size to 3,000. In \cite{chang23b_interspeech}, they set the BPE size to 6,000, achieving greater length reduction.} through the use of de-duplication and BPE subword-modeling. As we know, the computational cost of the normal attention mechanism~\cite{NIPS2017_3f5ee243} is quadratic relative to the input sequence length. Therefore, employing discrete representations for ASR training effectively reduces the normal attention mechanism's cost to one-quarter, thereby achieving considerable efficiency. 

\begin{table}[t]
    \centering
    \caption{Average input sequence length of Train \& Dev set of (1) Original Sequence (2) Sequence after De-duplication (DD) (3) Sequence after De-duplication (DD) + BPE. The number in brackets represents the reduced length percentage.} 
\begin{tabular}{l|c|c|c}
\toprule
\textbf{Splits} & \textbf{Original} & \textbf{DD} & \textbf{DD + BPE}  \\ \midrule
Train            & 393.26            & 272.31 (31\%)     & 202.92 (48\%)  \\ \midrule
Dev              & 333.01            & 226.30 (32\%)     & 180.08 (46\%)  \\ 
\bottomrule
\end{tabular}

    \label{tab:sequence}
\end{table}

\subsection{Benefit summarization}
\label{sec:benefit}
In this section, we compare and summarize the advantages of using continuous and discrete representation ASR based on Section~\ref{subsec:statistics}. Here's a detailed exploration of the benefits of each approach:

    \noindent \textbf{Continuous Representation. } While it offers strong performance, it incurs significant storage and I/O costs. Additionally, its sequence length cannot be easily reduced, leading to higher training and inference costs due to longer input sequences.
    
    \noindent \textbf{Discrete Representation. } This approach benefits from a smaller storage size and faster I/O during training and transmission. Its input sequence length can be reduced, which in turn reduces inference costs. However, it usually shows poor performance.
   

The comparison of our proposed fusion mechanism and self-augmented representations will be presented in Section~\ref{subsec:analysis}. Most importantly, \textbf{our proposed methods preserve all of the advantages of discrete representation} but with enhanced performance.

\section{Methodologies}
In this section, we first introduce the general picture and details of our fusion mechanism (Sec.~\ref{sec:fusion}).
Next, we introduce two representation augmentation methods to derive self-augmented discrete representations (Sec.~\ref{sec:repr_tfm}).

\subsection{Discrete representation fusion mechanism}
\label{sec:fusion}

\begin{figure}[t]
    \centering
    \includegraphics[width=0.6\linewidth]{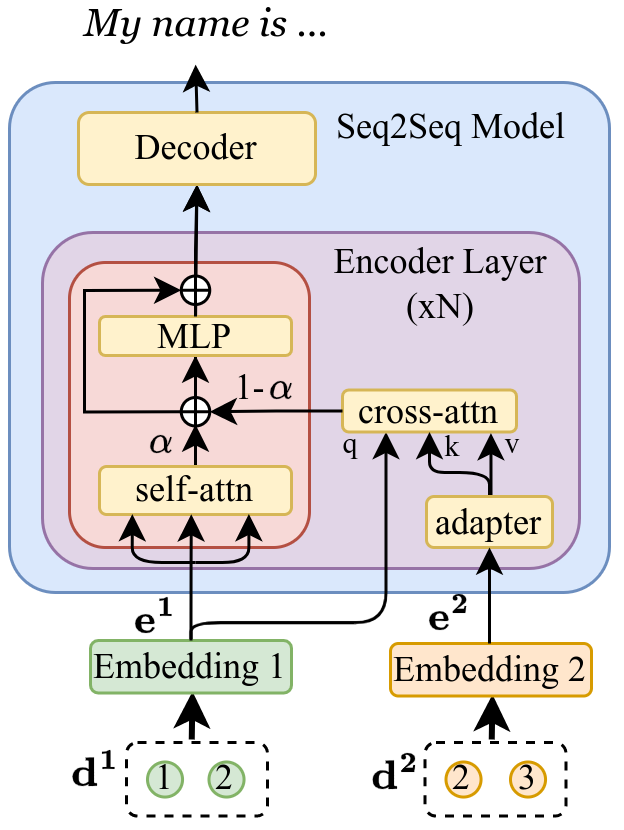}
    \caption{Proposed fusion mechanism. We fuse primary and secondary representations ($\mathbf{d^1}$ \& $\mathbf{d^2}$) to perform end-to-end ASR training.}
    
    \label{fig:fuse}
\end{figure}

As discussed in Section~\ref{sec: intro}, due to the non-linear misalignment between representations, we cannot simply concatenate them along the last dimension. Additionally, concatenating along the time dimension would result in excessively lengthy input sequences. To address these challenges, we introduce our fusion mechanism for discrete representations. Figure~\ref{fig:fuse} provides a visual overview of this mechanism, which primarily utilizes the cross-attention mechanism to address the misalignment between representations. 
The core component of our fusion approach is highlighted in the purple block of Figure~\ref{fig:fuse}. Differing from the standard encoder layer, which typically comprises a self-attention layer and a multi-layer perceptron (MLP) layer, our fusion mechanism incorporates an additional cross-attention layer to merge the discrete representations.

In Figure~\ref{fig:fuse}, the two discrete representations $\mathbf{d^1}$ and $\mathbf{d^2}$ are named as primary and secondary discrete representation, where $\mathbf{d^{i}} = [d_{1}^{i}, d_{2}^{i} ..., d_{T_{i}}^{i}]$, $i \in \{1, 2\}$.
Here, $d_{t}^{i}$ is the $t$\textsuperscript{th} discrete unit of $\mathbf{d^{i}}$, and $T_{i}$ is the length of $\mathbf{d^{i}}$. 
First, $\mathbf{d^1}$ and $\mathbf{d^2}$ go through their respective embedding layers to get $\mathbf{e^{1}}$ and $\mathbf{e^{2}}$, where $\mathbf{e^{i}} = [e_{1}^{i}, e_{2}^{i}, ..., e_{T_{i}}^{i}], e_{t}^{i} \in \mathbb{R}^{\text{D}_\text{emb}}, i \in \{1, 2\}$.
Here $\text{D}_\text{emb}$ is the dimension of two embedding layers. 
Next, $\mathbf{e^{1}}$ is subjected to self-attention within a typical encoder framework. Concurrently, it interacts with $\mathbf{e^{2}}$ in a cross-attention layer where $\mathbf{e^{1}}$ serves as the query and $\mathbf{e^{2}}$ as both key and value. This interaction results in $\mathbf{e^{f1_{\text{ca}}}}$. The purpose of this cross-attention at each encoder layer is to deeply integrate the information from the secondary representation into the model.
Last, we use a weighted-sum mechanism to combine the self-attention output $\mathbf{e^{f1_\text{sa}}}$ and cross-attention output $\mathbf{e^{f1_\text{ca}}}$ to get $\mathbf{e^{fuse}}$. This weighted-sum mechanism acts as a learnable gate, regulating the information flow between the primary and secondary representations. Note that $\mathbf{e^{2}}$ go through an adapter \footnote{each encoder layer has different cross-attention layers and adapters.} before entering the cross-attention layer. 

To be more specific, we send $\mathbf{e^{1}}$ into the encoder. $\mathbf{e^{1}}$ first go through self-attention layer to get $\mathbf{e^{f1_\text{sa}}}$ where $\mathbf{e^{f1_\text{sa}}} = [e_{1}^{f1_\text{sa}}, ..., e_{T_{1}}^{f1_\text{sa}}],$ $e_{t}^{f1_\text{sa}} \in \mathbb{R}^{\text{D}_\text{emb}}$. That is:
\begin{equation}
    \mathbf{e^{f1_\text{sa}}} = \textrm{Attention}(Q=K=V=\mathbf{e^{1}})  
\end{equation}
Also, we fuse $\mathbf{e^{1}}$ and $\mathbf{e^{2}}$ to get $\mathbf{e^{f1_\text{ca}}} =[e_{1}^{f1_\text{ca}}, ..., e_{T_{1}}^{f1_\text{ca}}]$, $e_{t}^{f1_\text{ca}} \in \mathbb{R}^{\text{D}_\text{emb}}$, with the following operation:
\begin{equation}
    \mathbf{e^{f1_\text{ca}}} = \textrm{Attention}(Q=\mathbf{e^{1}}, K=V=\textrm{Adapter}(\mathbf{e^{2}}))  
\end{equation}
where Adapter is a down-projection linear layer followed by a non-linear activation and an up-projection linear layer.

Obtaining $\mathbf{e^{f1_\text{sa}}}$ and $\mathbf{e^{f1_\text{ca}}}$, we apply weighted-sum to combine them to get $\mathbf{e^{fuse}}$. That is:
\begin{equation}
    \mathbf{e^{fuse}} = \alpha \cdot \mathbf{e^{f1_\text{sa}}} + (1 - \alpha) \cdot \mathbf{e^{f1_\text{ca}}} 
\end{equation}
where $\alpha$ is a learnable parameter. 

Last, we direct the $\mathbf{e^{fuse}}$ to the MLP module with a residual path as a normal encoder layer. It's important to mention that our fusion mechanism is scalable, allowing for the integration of more than two representations by adding additional cross-attention layers and expanding the weighted-sum component in each encoder layer. For demonstration purposes, in this paper, we have limited our focus to fusing only two discrete representations.
\subsection{Self-augmented representations}
\label{sec:repr_tfm}
The most common way to employ our fusion mechanism is to derive the required discrete representations from two SSL models. However, in scenarios where resources are extremely limited or inference cost is critical, the forward pass of two SSL models may be infeasible for the ASR system. To address these challenges, we propose to employ fusion with \textbf{self-augmented representations}, which can be derived by performing simple and efficient transformations on a single continuous SSL representation. In other words, with the proposed self-augmentation, we can derive two discrete representations with only one SSL model forward pass. In summary, the benefits of adopting self-augmented discrete representations include: (1) eliminating the dependency of our proposed fusion mechanism on the second SSL model; and (2) reducing inference costs. 

In this section, we introduce two approaches to augment continuous representations to derive self-augmented discrete representations with complementary information, which benefits fusion performance (see Self-Augment in Figure~\ref{fig:sub1}).

\noindent\textbf{(I) Reshape}.
A sequence of continuous representations is typically represented as a 2D array with dimensions $(T, \text{D}_\text{ssl})$. Instead of directly discretizing each frame representation in this standard format, we employ a simple reshape technique on the sequence before discretization, adjusting the dimensions from $(T, \text{D}_\text{ssl})$ to $(2T, \frac{\text{D}_\text{ssl}}{2})$. This involves splitting each frame ($h_i$) into two parts: the first part contains the initial $\frac{\text{D}_\text{ssl}}{2}$ dimensions, and the second part contains the remaining dimensions. We then carry out the discretization process, which adjusts the shape to $(2T, )$.

This method of augmentation allows us to represent each frame with two discrete units instead of one. By doing so, it provides a more fine-grained temporal representation of the information within each frame, potentially enhancing the detail of the representation. Although this method doubles the sequence length, employing it primarily as the secondary representation mitigates the potential increase in computational cost, maintaining a manageable sequence length within the model.


\noindent\textbf{(II) Delta}.
In Mel-frequency cepstral coefficients (MFCC), dynamic features such as first- (delta) and second-order (delta-delta) frame-wise differences are widely used.
Motivated by this, we also derive delta features from continuous representations and then perform discretization.
Different from MFCC, we discretize the delta features instead of concatenating them with the original features.
Although the temporal resolution remains unchanged, discretized delta representations could explicitly provide temporally dynamic information.

\section{Experimental Settings}
\label{sec:experiment}
\subsection{Dataset}
\label{subsec:dataset}
We evaluated the proposed method on LibriSpeech-100h~\cite{libirspeech} and ML-SUPERB~\cite{shi2023mlsuperb}.
The setting is aligned with the Discrete Speech Unit Challenge\footnote{\scriptsize{\url{https://www.wavlab.org/activities/2024/Interspeech2024-Discrete-Speech-Unit-Challenge/}}\label{fn:website}}.
LibriSpeech-100h evaluates the English ASR capability, providing 100 hours of clean English paired data.
On the other hand, ML-SUPERB assesses the multilingual ASR capability, consisting of approximately 200 hours of data from 143 languages.
For training, we mixed the training sets from LibriSpeech and ML-SUPERB and then performed multi-lingual training.
For evaluation, we evaluated our model on the \{dev-clean, dev-other, test-clean, test-other\} sets of LibriSpeech and the test set of ML-SUPERB.

\subsection{Model configuration \& discretization}
\label{subsec:model_config}
\noindent\textbf{Feature extractor}.
We adopted WavLM-Large~\cite{Chen2021WavLMLS} and MMS-1B~\cite{pratap2023mms} as the feature extractors. We use the output from layer 21 for WavLM-Large and layer 48 for MMS-1B. 
Both of them are SSL models, but WavLM is pre-trained in English data with denoising pre-training tasks, while MMS-1B is pre-trained in multilingual data with 1162 languages. We chose these models to capture complementary information from their respective representations, which is anticipated to enhance the performance of our fusion mechanism.

\noindent\textbf{Discretization}.
As stated by ~\cite{chang2023exploring}, clustering-based methods show inherent versatility for different tasks. Therefore, we follow Chang \emph{et al.}~\cite{chang2023exploring} to use K-Means clustering to discretize the continuous representations. 
The number of clusters is set to 2,000 for all models. To save computational cost, we only use 30\% of the training set data to train the K-Means model.
 
\noindent\textbf{De-duplication \& subword modeling}.
Again, following Chang \emph{et al.}~\cite{chang2023exploring}, we performed de-duplication and subword modeling (see Figure~\ref{fig:discretization}) to further reduce the length of discrete representation. For the BPE subword modeling, the BPE size is set to 3,000. 

\subsection{Implementation details}
\label{subsec:details}
We use the ESPnet~\cite{watanabe2018espnet} toolkit to run all the experiments.
For the delta feature calculation, we used the Librosa~\cite{mcfee2015librosa} implementation by setting the window width to 9.
For the ASR model, we adopted a Transformer~\cite{NIPS2017_3f5ee243}-based encoder-decoder architecture.
The dimension of the embedding layer $\text{D}_\text{emb}$ was set to 512.
We added a linear layer to reduce the dimension from 512 to 256 after the embedding layer.
The encoder and decoder consisted of 12 and 6 layers, respectively.
In both self- and cross-attention modules, the number of heads was set to 4, and the linear units were set to 1024, with a dropout probability of 0.1.
For MLP layers, the linear units were set to 1,024.
For the adapter, we implemented a design consisting of a down-projection linear layer, a non-linear activation, and an up-projection linear layer, where the bottleneck size was 128.
During the training, we utilized the Adam~\cite{Kingma2014AdamAM} optimizer, setting the learning rate at 0.0005, applying a minimal weight decay of 1e-6, and incorporating 5000 warm-up steps. 
We performed joint attention-based encoder-decoder and connectionist temporal classification (CTC) training, with a CTC weight of 0.3.
For the inference phase, we obtained the ASR output using beam search with a beam size of 20.
Besides, we did not use additional language models.

\subsection{Fusion variants \& baselines}
\label{subsec:baseline}

\noindent \textbf{Fusion variants.}
We explored three distinct combinations of discrete representations.
The primary representation ($\mathbf{d^1}$) was consistently set as the MMS-1B across all scenarios.
For the secondary representation ($\mathbf{d^2}$), we employed three different variants: (1) WavLM-Large, (2) the reshaped representation of MMS-1B, and (3) the delta representation of MMS-1B as introduced in Section~\ref{sec:repr_tfm}. For the order of primary and secondary representations, we have an analysis in Section~\ref{subsec:analysis}.

For a fair comparison, we established several baselines under the same experimental settings as described in Section~\ref{subsec:details}:

\noindent \textbf{Continuous representation.} For this baseline, we utilized the continuous representation from the MMS-1B. In line with our experimental settings, we implemented a linear layer to reduce the dimension from 1280 to 80 and added a convolution layer to halve the sequence length. Also, we remove the down-sample linear
layer after the embedding layer~\footnote{Encoder's output dimension will be 512.}. The rest of the model configuration is aligned with that in Section~\ref{subsec:model_config}. This setup is considered a strong baseline, primarily due to its higher storage and computational demands, as discussed in Section~\ref{sec:benefit}. Its bitrate is calculated at footnote~\ref{fn:2M}

 \noindent \textbf{Concatenated discrete representations.} As introduced in Section~\ref{sec:fusion}, a simple method for fusing two discrete representations is concatenating them along the time dimension. However, this method extends the input sequence in training and inference. We tested with concatenated discrete representations from WavLM-Large and MMS-1B.

 \noindent \textbf{Non-Fusion discrete representation.} In this baseline, we trained the model using only one discrete representation, as conducted in \cite{chang23b_interspeech}. We tested discrete representations from WavLM-Large and MMS-1B separately.
    
 \noindent \textbf{Non-Fusion high bitrate discrete representation.} Given that our fusion mechanism employs two discrete representations, resulting in a doubled required bitrate, we provided a non-fusion baseline with a comparable bitrate using the non-fusion discrete representation MMS-1B baseline but without de-duplication and BPE.



\subsection{Evaluation Metrics}
\label{sec:metric}
In alignment with the Discrete Speech Unit Challenge, our evaluations focus on two key metrics: \textbf{character error rate (CER)} and \textbf{bitrate}, for all experiments. For LibriSpeech-100h~\cite{libirspeech}, we calculate the micro-average CER across the four evaluation sets within LibriSpeech-100h. For ML-SUPERB~\cite{shi2023mlsuperb}, we continue to use CER, adhering to ML-SUPERB's established metric for evaluation.

To further evaluate the efficiency of transmission, we also measure the bitrate for all test sets, including both LibriSpeech-100h and ML-SUPERB. We denote the discrete representations as $\{\mathbf{d^{1}}, \ldots, \mathbf{d^{M}}\}$, where $\mathbf{d^{i}}$ represents the $i^{th}$ stream of discrete representations, and $\mathbf{M}$ denotes the total number of streams. The vocabulary size for the $i^{th}$ stream is expressed as $V_i$, and the length of the $i^{th}$ stream is noted as $T_i$. Given that the total length of the test sets is $N$ (in seconds), we define the bitrate using the formula:

\begin{equation}
\text{Bitrate} = \sum_{i=1}^{\mathbf{M}} \left( \frac{T_i}{N} * \log_2{(V_i)} \right)
\end{equation}

The formula adheres to the definition on the challenge website\footnotemark[\value{footnote}].

\section{Result \& Analysis}
\subsection{Quantitative result}
\label{sec:quantitative_result}
\begin{table}[t]
    \caption{ASR Results of our fusion variants and baselines (see Section~\ref{subsec:baseline}). $\mathbf{d^1}$ is the primary representation, and $\mathbf{d^2}$ is the secondary representation. Rows with a missing secondary representation (-) represent non-fusion baselines. We report the CER for both the LibriSpeech (LS) averaged dev/test set and the ML-SUPERB (MS) test set. For the bitrate, we report the number based on the Section~\ref{sec:metric}. }
    \centering
{\begin{threeparttable}
\centering
\resizebox{\linewidth}{!}{
\begin{tabular}{l|l|c|c|c}
\toprule
\multicolumn{1}{c|}{\textbf{$\mathbf{d^1}$}}                      & \multicolumn{1}{c|}{\textbf{$\mathbf{d^2}$}}                                                              & \textbf{LS$\downarrow$}                        & \textbf{MS$\downarrow$}                        & \textbf{Bitrate $\downarrow$}                  \\ \toprule \midrule
\begin{tabular}[c|]{@{}l@{}}Continuous\\ MMS-1B\end{tabular}                                                 & \multicolumn{1}{c|}{-}                                                               &        2.34               &    10.89                  & 2048000              \\\midrule \midrule 
\begin{tabular}[c]{@{}l@{}}WavLM-Large\end{tabular} & \multicolumn{1}{c|}{-}                                                               & 2.37                      & 22.4                      & 356.19                   \\ \midrule
MMS-1B                                                 & \multicolumn{1}{c|}{-}                                                               & 2.32                      & 14.32                     & \textbf{280.86}                   \\ \midrule
\begin{tabular}[c|]{@{}l@{}}High Bitrate\\ MMS-1B\end{tabular}                                                 & \multicolumn{1}{c|}{-}                                                               &      2.52                 &      14.38                & 556.15                   \\\midrule

 Concat MMS-1B                                                 & \begin{tabular}[c|]{@{}l@{}}WavLM-Large\end{tabular}          &        1.92                &   11.25                    & 665.13            \\ \midrule  \midrule
MMS-1B                                                 & \begin{tabular}[c|]{@{}l@{}}WavLM-Large\end{tabular}          & \textbf{1.89}                       & \textbf{10.87}                      & 665.13            \\ \midrule
MMS-1B                                                 & \begin{tabular}[c|]{@{}l@{}}MMS-1B\\ Reshape\end{tabular} & 2.26                      & 12.22                     & 1024.90            \\ \midrule
MMS-1B                                                 & \begin{tabular}[c|]{@{}l@{}}MMS-1B\\ Delta\end{tabular}          & 2.17                      & 11.69                     & 648.52            \\ \bottomrule
\end{tabular}
}
\end{threeparttable}}

    \label{tab: Main_Result}
\end{table}

\noindent\textbf{ASR results}.
Table~\ref{tab: Main_Result} showcases the CER of our baselines and fusion variants on the LibriSpeech and ML-SUPERB datasets. Rows without a secondary representation (-) indicate non-fusion baselines.

The results demonstrate that all fusion variants consistently outperform the non-fusion baselines, even with comparable bitrates. This underscores the effectiveness of our proposed fusion mechanism. Remarkably, fusion variants achieve performance slightly superior to the continuous representation baseline while using only 0.3\% of its bitrate.
In particular, the fusion of MMS-1B with WavLM-Large showcases the highest gains, achieving a 19\% and 24\% relative improvement on LibriSpeech and ML-SUPERB. This enhancement across both English-only and multilingual datasets illustrates that our fusion mechanism adeptly integrates complementary information from each representation. Even compared to the concatenated baseline, our approach exhibits better performance without the drawback of extended input sequences. Additionally, our self-augmented discrete representations have also demonstrated improvements over the non-fusion MMS-1B baseline.

From a computational standpoint, our fusion mechanism efficiently uses multiple discrete representations, resulting in only a 24\% increase in the number of parameters. By leveraging cross-attention, we avoid doubling the input sequence length, which helps prevent a quadratic increase in computational requirements. Moreover, since the extra computation is confined to the cross-attention layers within the encoder, it minimally impacts the overall training and inference times.

\noindent\textbf{Bitrate}.
Table \ref{tab: Main_Result} also presents the bitrate for each baseline and our fusion variants. While fusing additional discrete representations naturally increases the bitrate, the application of de-duplication and subword modeling techniques ensures that the overall bitrate remains within an acceptable range compared to the continuous representation baseline. Overall, our fusion mechanism enhances the performance of discrete representations for ASR without incurring significant additional transmission costs.

\subsection{Qualitative analysis}
\label{subsec:analysis}
\noindent\textbf{Fusion mechanism \& self-augmented representations benefit}.
In Table~\ref{tab:benefit}, we summarize the benefits of using both continuous and discrete representations for ASR~\ref{sec:benefit}, integrating our proposed methods into the discussion. Our proposed fusion mechanism (see Section~\ref{sec:fusion}) enhances the model's performance and preserves all the benefits of discrete representations—despite doubling the numbers in Table~\ref{tab:storage}, storage costs remain low, and I/O efficiency is preserved. Importantly, this mechanism does not lead to longer input sequences.

Moreover, our innovative self-augmented representations (see Section~\ref{sec:repr_tfm}) can reduce the inference costs. The effectiveness and benefits of these approaches are comprehensively compared in Table~\ref{tab:benefit}, highlighting the improvements across various aspects.
\begin{table}[]
    \centering
    \caption{Comparison across different representation types. An \textbf{O} indicates that the representation type has the corresponding advantage, an \textbf{X} indicates the absence of the advantage, and a \textbf{$\Delta$} indicates mediocre performance.}
    \resizebox{\linewidth}{!}{
\begin{tabular}{l|c|c|c|c} 
\toprule
\textbf{Representation}  & \textbf{Perfor-} & \textbf{Storage \&} & \textbf{Reduced Seq-} & \textbf{Low Infe-} \\
\textbf{Type} & \textbf{mance} & \textbf{Faster I/O} & \textbf{uence Length} & \textbf{rence Cost} \\
\midrule
Continuous & O & X & X & $\Delta$ \\\midrule
Discrete & X & O & O & O \\ \midrule
Discrete Fusion & O & O & O & $\Delta$ \\ \midrule
+ self-augmented & O & O & O & O \\
\bottomrule
\end{tabular}
}

    \label{tab:benefit}
    \vspace{-3mm}
\end{table}

\noindent\textbf{Primary \& secondary representations}.
\begin{table}[t]
    
    \centering
    \caption{Ablation for primary representation determination.}
{\begin{threeparttable}
\centering
\resizebox{.95\linewidth}{!}{
\begin{tabular}{l|l|c|c|c}
\toprule
\multicolumn{1}{c|}{$\mathbf{d^1}$}                      & \multicolumn{1}{c|}{$\mathbf{d^2}$}                                                              & \textbf{LS$\downarrow$}                        & \textbf{MS$\downarrow$}                        & \textbf{Bitrate $\downarrow$}                  \\ \toprule \midrule
MMS-1B                                                 & \begin{tabular}[c|]{@{}l@{}}WavLM-Large\end{tabular}          & \textbf{1.89}                       & \textbf{10.87}                      & 665.13            \\ \midrule
 \begin{tabular}[c|]{@{}l@{}}WavLM-Large\end{tabular}                                            & MMS-1B          & 1.95                       & 11.22                      & 665.13                       \\ \bottomrule
\end{tabular}
}
\end{threeparttable}}

    \label{tab: ablation}
    \vspace{-3mm}
\end{table}
We investigated the impact of the order of primary and secondary representations.
For simplicity, we used MMS-1B and WavLM-Large as they showed the best performance in Table~\ref{tab: Main_Result}, where MMS-1B was used as the primary representation and WavLM-Large as the secondary one.
We then reversed their roles, making WavLM-Large primary and MMS-1B secondary.
Table~\ref{tab: ablation} shows that both configurations deliver effective performance. However, MMS-1B, when used as the primary representation, performs slightly better. Thus, we recommend using the representation with superior performance as the primary one. Additionally, this finding indicates that the weighted-sum mechanism skillfully controls the information flow between the two representations, maintaining robust performance irrespective of their order.
\noindent\textbf{Inference cost analysis}.
As outlined in Section~\ref{sec:repr_tfm}, utilizing our self-augmented representations significantly reduces inference costs. To quantify this reduction, we conducted experiments on the LibriSpeech dev-clean set, focusing on the time required for feature extraction. We compared the average times for two processes: (1) extracting continuous representations from raw waveforms using WavLM-Large, and (2) performing a delta transformation on the continuous representations from MMS-1B using only a CPU. This setup corresponds to our best-performing fusion variant (MMS-1B + WavLM-Large vs. MMS-1B + MMS-1B Delta) in Table~\ref{tab: Main_Result}.

This experiment utilized an RTX3090 GPU and an Intel Xeon Gold 6326 CPU, spanning 2703 raw waveform files. The results, including average times and standard deviations (STD), are as follows: WavLM-Large feature extraction took \textbf{0.72 $\pm$ 0.24 seconds}, while delta transformations required only \textbf{0.10 $\pm$ 0.01 seconds}. These statistics show that the time needed for delta transformations is just \textbf{13.8\%} of that required for a second SSL model's feature extraction, clearly highlighting the considerable decrease in inference costs.

\noindent\textbf{Language-wise analysis}.
In this study, we evaluate the language robustness of various secondary representations when combined with MMS-1B as the primary representation. We investigate WavLM-Large, Reshape, and Delta representations using results from the ML-SUPERB test set, with non-fusion MMS-1B serving as the baseline.
To quantitatively analyze language robustness across each fusion variant, we define the relative CER difference \text{$\mathrm{CER}^{l}_{\text{diff}}$} for each language $l$ as:
\begin{equation}
    \mathrm{CER}^{l}_{\text{diff}} = \frac{\mathrm{CER}^{l}_f - \mathrm{CER}^{l}_b}{\mathrm{CER}^{l}_b}
\end{equation}
where \text{$\mathrm{CER}^{l}_f$} and \text{$\mathrm{CER}^{l}_b$} represent the CER for the fusion variant and the baseline, respectively, on language $l$. This metric indicates the performance difference between the fusion variant and the baseline specific to language $l$. By computing the STD of \text{$\mathrm{CER}^{l}_{\text{diff}}$} across all languages, we can measure the variability of performance improvements or declines.
Further, to provide a clear understanding of language robustness, we categorize each \text{$\mathrm{CER}^{l}_{\text{diff}}$} into three groups: \textit{Decline}, \textit{Comparable} (where \text{$\mathrm{CER}^{l}_{\text{diff}}$} falls within $\pm$5\%), and \textit{Improved}. A robust fusion variant is characterized by a low \text{$\mathrm{CER}^{l}_{\text{diff}}$} STD, a higher count of \textit{Improved} languages, and fewer \textit{Decline} languages. The numbers are summarized in Table~\ref{tab:language}.
 
\begin{table}[t]
    \centering
    \caption{Counts of \textit{Decline}, \textit{Comparable}, and \textit{Improved} languages of fusion variants compared with the baseline. }


\begin{threeparttable}
\resizebox{0.85\linewidth}{!}{
\begin{tabular}{|l|c|c|c|}
\toprule
\begin{tabular}[c|]{@{}l@{}}\textbf{Secondary} \\ \textbf{Represention} \end{tabular}& \textit{Decline $\downarrow$} & \textit{Comparable} & \textit{Improved $\uparrow$} \\ \midrule
WavLM-Large & 5 & 31 & 106 \\ \midrule
Delta & \textbf{1} & 31 & 110 \\ \midrule
Reshape & 5 & 16 & \textbf{111} \\ \midrule
\bottomrule
\end{tabular}}
\end{threeparttable}
    
    \label{tab:language}
    \vspace{-3mm}
\end{table}
In our analysis, the STD of the \text{$\mathrm{CER}^{l}_{\text{diff}}$} for WavLM-Large, Reshape, and Delta representations are \textbf{0.19}, \textbf{0.14}, and \textbf{0.14}, respectively. This indicates that fusion with WavLM-Large is the least robust among the options. Furthermore, Table~\ref{tab:language} reveals that fusion with Delta representations resulted in the fewest \textit{Decline} languages, while Reshape had the highest number of \textit{Improved} languages. Surprisingly, despite WavLM-Large's generally lower CER, it shares the same number of \textit{Decline} languages as Reshape. These statistics suggest that self-augmented representations, especially Delta, are robust and demonstrate greater language independence. This is likely due to WavLM-Large being pre-trained primarily on English data, which may introduce language-specific biases.

Overall, while fusion with WavLM-Large often leads to superior performance, its effectiveness varies greatly by language. Conversely, self-augmented representations deliver more consistent improvements across languages, establishing them as language-agnostic options for fusion.

    

\vspace{-2mm}
    
\section{Conclusion}
\label{sec:conclusion}
\vspace{-2mm}
We propose a novel fusion mechanism that integrates two non-linearly misaligned discrete representations by utilizing an attention-based approach. Most importantly, our proposed methods preserve all of the advantages of discrete representation but with enhanced performance. Additionally, we developed ``self-augmented'' representations by efficiently transforming a single continuous SSL representation, enhancing our fusion mechanism's applicability by eliminating the need for the second SSL model and reducing inference costs. Our analysis demonstrates that fusing with these self-augmented representations is more language-robust than fusing with WavLM-Large discrete representations.
Experiments conducted on the LibriSpeech and ML-SUPERB datasets have validated the effectiveness of the proposed approaches, yielding up to 19\% and 24\% relative CER improvements compared with the non-fusion MMS-1B baseline, respectively. Most importantly, our proposed approaches achieve performance slightly
superior to the continuous representation baseline using only 0.3\% of its bitrate.


\end{document}